\documentclass[letter]{aa} 
\usepackage{graphicx}
\usepackage{txfonts}
\usepackage{textcomp}
\usepackage{natbib}
\usepackage{color}
\usepackage[breaklinks=false]{hyperref}
\newcommand{\Ms}{{\ensuremath{\mathrm{M}_{\odot}}}}
\newcommand{\Rs}{{\ensuremath{\mathrm{R}_{\odot}}}}
\newcommand{\Mpy}{\Ms\,{\rm yr}{\ensuremath{^{-1}}}}

\newcommand{\dMac}{\ensuremath{\dot M_+}}
\newcommand{\dMwnd}{\ensuremath{\dot M_-}}
\newcommand{\dJ}{\ensuremath{\dot J_{\rm eff}}}
\newcommand{\dJkep}{\ensuremath{\dot J_K}}
\newcommand{\dJac}{\ensuremath{\dot J_{\rm acc}}}
\newcommand{\dJmag}{\ensuremath{\dot J_{\rm ext}}}
\newcommand{\gva}{{\sc genec}}

\newcommand{\vinf}{{\ensuremath{\rm v_\infty}}}
\newcommand{\vesc}{{\ensuremath{\rm v_{esc}}}}
\newcommand{\OG}{{\ensuremath{\Omega\Gamma}}-limit}
\newcommand{\Ro}{{\ensuremath{{\rm Ro}}}}

\begin{document}

\title{Magnetic braking of supermassive stars through winds}
\titlerunning{Magnetic braking of supermassive stars through winds}

\author{L. Haemmerl\'e, G. Meynet}
\authorrunning{Haemmerl\'e \& Meynet}

\institute{D\'epartement d'Astronomie, Universit\'e de Gen\`eve, chemin des Maillettes 51, CH-1290 Versoix, Switzerland\label{inst1}}

\date{Received ; accepted }

 
\abstract
{Supermassive stars (SMSs) are candidates for being progenitors of supermassive quasars at high redshifts.
However, their formation process requires strong mechanisms that would be able to extract the angular momentum of the gas that the SMSs accrete.}
{We investigate under which conditions the magnetic coupling between an accreting SMS and its winds
can remove enough angular momentum for accretion to proceed from a Keplerian disc.}
{We numerically computed the rotational properties of accreting SMSs that rotate at the \OG\
and estimated the magnetic field that is required to maintain the rotation velocity at this limit
using prescriptions from magnetohydrodynamical simulations of stellar winds.}
{We find that a magnetic field of 10 kG at the stellar surface is required to satisfy the constraints on stellar rotation from the \OG.}
{Magnetic coupling between the envelope of SMSs and their winds could allow for SMS formation by accretion from a Keplerian disc,
provided the magnetic field is at the upper end of present-day observed stellar fields.
Such fields are consistent with primordial origins.
}
 
   \keywords{stars: massive -- stars: rotation -- stars: mass-loss -- stars: formation -- stars: magnetic field -- stars: Population III}
 
\maketitle
%

\section{Introduction}
\label{sec-intro}

Supermassive stars (SMSs) of $10^4-10^5$ \Ms\ are candidates for being progenitors of the supermassive black holes
that power the quasars that have recently been discovered at redshift $z\sim7$ \citep{mortlock2011,wu2015,banados2018,wang2018,woods2018}.
The black-hole masses inferred by the observations are as high as $10^9-10^{10}$ \Ms,
which implies mass-accretion rates of 0.1 -- 10 \Mpy\ by a simple timescale argument.
Accretion at such high rates is thought to occur in primordial haloes where H$_2$ molecules have been destroyed,
for instance by an external Lyman-Werner flux from a nearby starburst \citep{latif2013e,becerra2015,regan2017,becerra2018,smidt2018}.

The properties of Population III SMSs that accrete at the rates of atomically cooled haloes have been studied by several authors
\citep{hosokawa2013,sakurai2015,umeda2016,nakauchi2017,woods2017,haemmerle2018a,haemmerle2018b,surace2018}.
They are found to evolve as `red supergiant protostars' \citep{hosokawa2013}, following the Hayashi limit upwards as their mass grows by accretion.
Their effective temperature remains locked on $\sim5000$ K while their luminosity, which is nearly at the Eddington value, increases linearly with their mass.
During their main accretion phase, their structure is made of a convective core ($\sim10\%$ in mass) that is initially triggered by H-burning,
a convective envelope ($\sim1\%$ in mass) due to the low temperatures on the Hayashi limit,
and a large intermediate radiative region ($\sim90\%$ in mass) in between (Fig.~\ref{fig-sys}).
Without metal lines, the radiative mass-losses of Pop III SMSs are inefficient
because the corresponding winds do not reach the escape velocity \citep{nakauchi2017}.
On the other hand, these objects are pulsationally unstable,
and the estimates of the mechanical mass-losses through pulsations give $10^{-3}$ \Mpy\ \citep{hosokawa2013}.
These rates are several orders of magnitude lower than the accretion rate, so that pulsation instability does not prevent SMSs from forming by accretion,
but it indicates that they could have winds on large scales.
SMSs eventually collapse to a black hole through the general relativistic (GR) instability \citep{chandrasekhar1964}
when their mass reaches $2-3\times10^5$ \Ms\ \citep{umeda2016,woods2017,haemmerle2018a}.
Accretion continues after the collapse until the black-hole mass reaches $10^9-10^{10}$ \Ms\ at $z\sim7$.

The SMSs that form by accretion must be slow rotators, with surface velocities lower than about 10\% of their Keplerian velocity \citep{haemmerle2018b}.
This is a consequence of the \OG\ \citep{maeder2000}, which is relevant for stars that are close to the Eddington limit.
It requires the accreted angular momentum to be about 1\% Keplerian,
meaning that the 99\% of the angular momentum from a Keplerian accretion disc must be extracted from the accreted gas.
This is a particular case of the angular momentum problem that is general to star formation.
The commonly invoked mechanisms are magnetic fields, gravitational torques from spiral arms, or viscosity originating from other causes
\citep{hosokawa2016,takahashi2017,pandey2019}.

The aim of the present work is to investigate the role of a magnetic coupling between the star and its mechanical winds
in extracting the excess of angular momentum accreted from a Keplerian disc.
The magnetic properties of SMSs are not known.
Here, we estimate the intensity of the magnetic field that is required at the stellar surface
for the star to be maintained under the \OG\ during the accretion process.
\cite{haemmerle2018b} computed models of SMSs that accreted angular momentum at a constant fraction of the Keplerian angular momentum.
We do not constrain the angular momentum accretion rate here
but instead assume a configuration that rotates at a maximum rate at each time step by directly constraining the surface velocity.
This allows us to quantify more precisely the constraint on the accretion of angular momentum at the various evolutionary stages.
The method is described in Sect.~\ref{sec-method}.
The results are given in Sect.~\ref{sec-result} and are discussed in Sect.~\ref{sec-discuss}.
We conclude in Sect.~\ref{sec-out}.

\section{Method}
\label{sec-method}

\subsection{Description of the system}
\label{sec-sys}

\begin{figure*}\begin{center}
\includegraphics[width=0.55\textwidth]{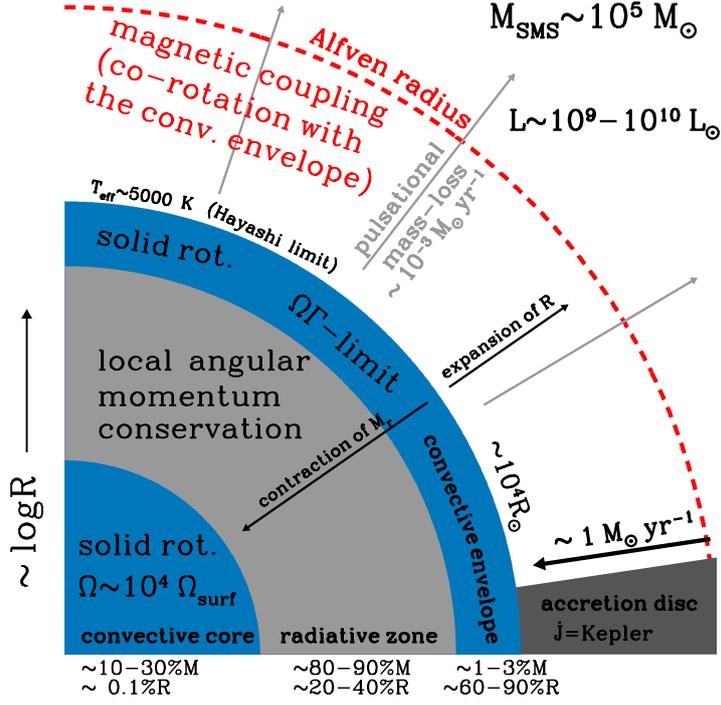}
\caption{Schematic view of the physical system (see Sect.~\ref{sec-sys} for a description).}
\label{fig-sys}
\end{center}\end{figure*}

A schematic picture of the system is shown in Fig.~\ref{fig-sys}.
The star consists of a convective core and a convective envelope (shown in blue), with a radiative region in between (shown in light grey).
It accretes through a Keplerian disc at $\sim1$ \Mpy\ and looses mass through pulsation instability at $\sim0.001$ \Mpy.
The convective envelope and the winds co-rotate as a solid body at the \OG\ (Eq.~\ref{eq-omgam}) up to the Alfv\'en radius.
The photospheric radius increases with time.
In the stellar interior, the layers contract from the convective envelope to the convective core.
They maintain their specific angular momentum while crossing the radiative zone, and eventually deliver it to the core, which rotates as a solid body.
In the next two subsections, we describe the properties of the star and the magnetic coupling in more detail.

\subsection{Stellar model}
\label{sec-star1}

The internal structure of SMSs that form by accretion is not affected by rotation because their velocities are slow \citep{haemmerle2018b}.
Moreover, their short lifetime ($\sim10^5$ yr) means that mixing processes other than convection are inefficient in transporting angular momentum.
We can therefore assume local angular momentum conservation in radiative regions, and convective regions are expected to rotate as solid bodies.
This allows us to post-process rotation on the non-rotating models described in \cite{haemmerle2018a},
which are computed with the \gva\ stellar evolution code that includes accretion \citep{eggenberger2008,haemmerle2014,haemmerle2016a}.

We proceeded in the following way.
We assumed that the stellar surface rotates at the angular velocity given by the \OG\ \citep{maeder2000,haemmerle2018b},
\begin{equation}
\Omega_*^2={3\over2}\Omega_K^2(R)(1-\Gamma),
\label{eq-omgam}\end{equation}
where $\Omega_K(R)=\sqrt{GM/R^3}$ is the Keplerian velocity at the stellar surface ($M$ is the stellar mass and $R$ its photospheric radius)
and $\Gamma$ is the Eddington factor.
This expression is only relevant for $\Gamma\gtrsim0.6.$ 
For lower $\Gamma$, the Keplerian limit still holds, and we assumed $\Omega_*=\Omega_K(R)$ in this case.
Knowing the surface velocity, we computed the rotational structure of the star with two assumptions:
solid rotation in convective regions, and local angular momentum conservation in radiative regions.
The rotational properties of the convective envelope in solid rotation are fully determined by the surface velocity.
Rotation in the radiative region is computed by assuming that the layers enter in the upper part of the radiative zone
with the angular velocity of the surface and contract with constant specific angular momentum until they are incorporated in the convective core.
We used this specific angular momentum to compute the increase in the angular momentum of the convective core,
which determines its rotational properties because we assumed solid rotation.
Knowing the rotation profile of the star, we computed the effective rate of the increase \dJ\ in its total angular momentum.

We assumed that accretion proceeds at a rate \dMac\ through a Keplerian disc in the equatorial plan (Fig.~\ref{fig-sys}).
Thus the angular momentum advected inside the star by accretion is given by
\begin{equation}
\dJac=\dJkep=\dMac R^2\Omega_K(R)=\dMac\sqrt{GMR}.
\label{eq-jac}\end{equation}
From the effective increase \dJ\ of the angular momentum of the star given by the stellar model,
we estimated the angular momentum excess that must be extracted for the star to rotate at the \OG:
\begin{equation}
\dJmag=\dJac-\dJ=\dMac\sqrt{GMR}-\dJ.
\label{eq-jdif}\end{equation}

\subsection{Magnetic coupling with winds}
\label{sec-mag1}

The next step was to estimate the magnetic field that is required for loosing the angular momentum excess through coupling with wind.
We followed the prescriptions of \cite{uddoula2008,uddoula2009}, which are based on two-dimensional magnetohydrodynamical simulations of stellar winds.
According to these prescriptions, the magnetic coupling with the winds relies on the magnetic confinement parameter
\begin{equation}
\eta_*:={B_*^2R^2\over\dMwnd \vinf},
\label{eq-eta}\end{equation}
where $B_*$ is the magnetic field at the stellar surface (at the equator), \dMwnd\ is the mass-loss rate (absolute value),
and \vinf\ is the terminal wind velocity.
The terminal velocity is related empirically to the escape velocity \vesc\ at the stellar surface,
with a ratio that depends on the spectral type \citep{lamers1995}.
For stars at the Hayashi limit,
\begin{equation}
\vinf=0.72\,\vesc.
\label{eq-vinf}\end{equation}
The distance on which the coupling acts is given by the Alfv\'en radius, which is related to the confinement parameter by \citep{uddoula2009}
\begin{equation}
{r_A\over R}=0.29+(\eta_*+0.25)^{1/4}.
\label{eq-alf}\end{equation}
The loss of angular momentum due to the wind magnetic braking process is then given by
\begin{equation}
\dJmag={2\over3}\dMwnd r_A^2\Omega_*.
\label{eq-jmag}\end{equation}
From \dJmag\ of Eq.~(\ref{eq-jdif}), we can derive the Alfv\'en radius and magnetic field that are required for the SMS to rotate at the \OG\
using Eq.~(\ref{eq-eta} -- \ref{eq-jmag}).

\section{Results}
\label{sec-result}

\begin{figure}\begin{center}
\includegraphics[width=0.45\textwidth]{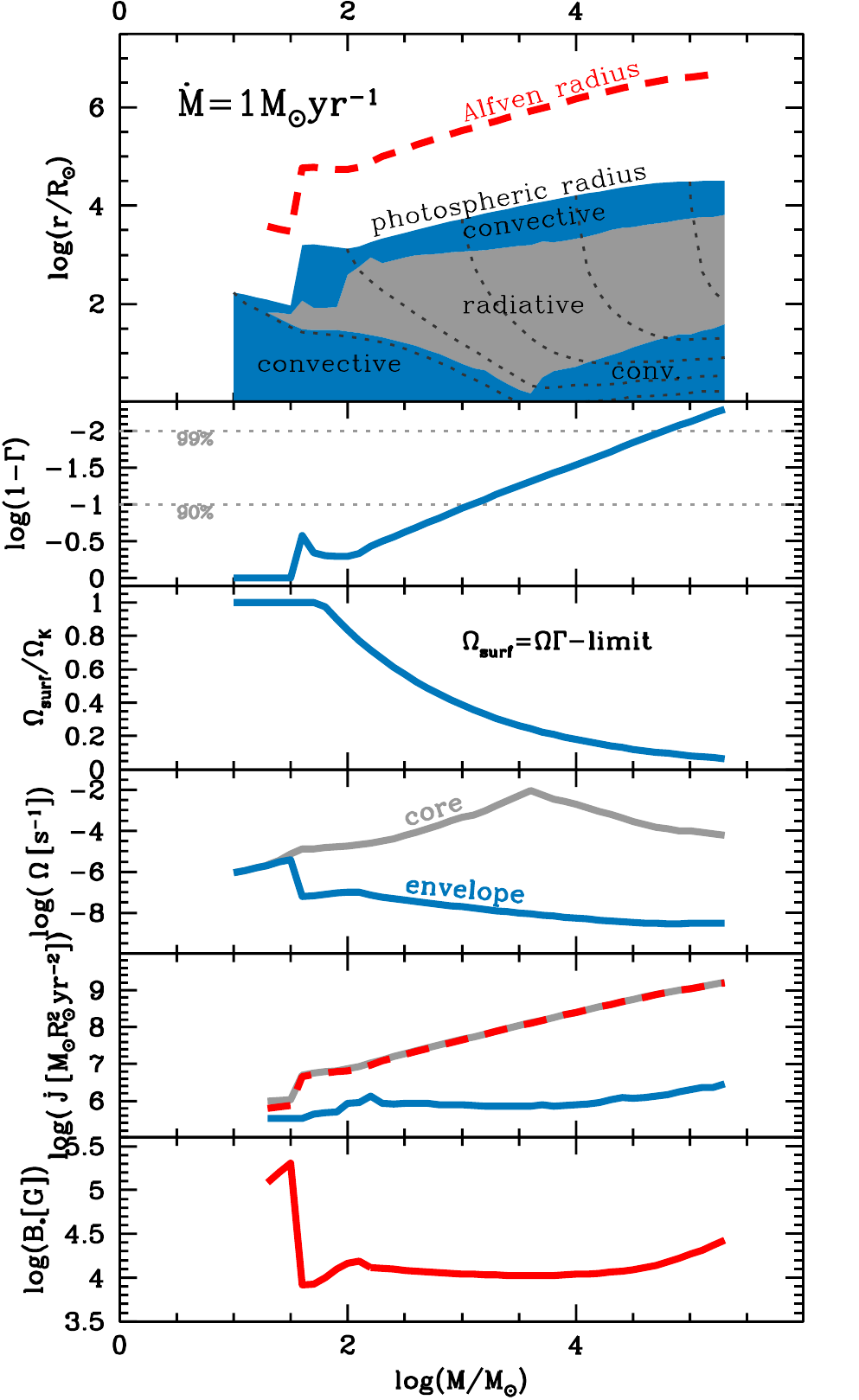}
\caption{Properties of the stellar model with $\dMac=1$~\Mpy\ rotating at the \OG
as a function of the stellar mass (time coordinate in case of constant accretion).
The upper panel shows the stellar structure and the Alfv\'en radius.
Blue areas are convective regions, and the grey area is the radiative zone.
The black dotted lines are Lagrangian layers (iso-mass) of $\log M_r/\Ms=$1, 2, 3, 4, and 5.
The second panel shows the Eddington factor in $\log(1-\Gamma)$.
Grey dotted lines indicate $\Gamma=0.9$ and 0.99.
The third panel shows the constraint we impose on the angular velocity at the surface (\OG, Eq.~\ref{eq-omgam}),
as a ratio to the Keplerian velocity.
The fourth panel shows the angular velocity of the envelope (in blue) and of the core (in grey).
The fifth panel shows the angular momentum accretion rate \dJac\ (Keplerian, in grey),
the time derivative of the stellar angular momentum \dJ\ (in blue),
and the required angular momentum-loss rate \dJmag\ (red dashed line) computed with Eq.~(\ref{eq-jdif}).
The last panel shows the magnetic field that is required at the stellar surface according to Eq.~(\ref{eq-eta} -- \ref{eq-jmag}).}
\label{fig-ud}
\end{center}\end{figure}

\subsection{Rotational properties of the star}
\label{sec-star2}

We considered as a fiducial case the rates $\dMac=1$ \Mpy\ and $\dMwnd=0.001$ \Mpy\ (see Sect.~\ref{sec-intro} or \citealt{hosokawa2013}).
The results obtained with the method of Sect.~\ref{sec-method} are shown in Fig.~\ref{fig-ud}.
The top panel shows the internal structure of the star from the non-rotating models of \cite{haemmerle2018a}.
In order to avoid transitions between convective and radiative transfer in the early phase, which reflect the arbitrary choice of the initial model,
we assumed that the initial 10 \Ms\ seed remains fully convective.
After the fully convective phase, we are left with a simple structure made of
a convective core, a convective envelope, and a radiative region in between.
There are two convective cores that originate from two different causes.
The first core ($M<4000$ \Ms) reflects the choice of the structure of the initial seed that contracts.
This feature is present in the models of \cite{haemmerle2018a}, but is simplified here.
The second core is driven by H-burning, which starts at $M\simeq4000$~\Ms.
The evolution stops at $\sim2\times10^5$ \Ms, when the star reaches the GR instability.
The Eddington factor is shown in the second panel.
After convergence at the Hayashi limit, it almost follows a power law in $M$, with values 0.9 -- 0.99 for $M>10^4$ \Ms.

The third panel of Fig.~\ref{fig-ud} shows the constraint we imposed on the surface velocity, that is, the \OG\ given by Eq.~(\ref{eq-omgam}).
The angular velocities of the core and the envelope resulting from this constraint are shown in the fourth panel.
The initial model being fully convective, both velocities are initially identical, about $\Omega\simeq10^{-6}$~s$^{-1}$.
When the radiative region forms ($M\simeq20$~\Ms), the curves start to diverge.
The photosphere moves to the Hayashi limit at $M\simeq30$~\Ms,
and the surface velocity drops to $10^{-7}-10^{-8}$~s$^{-1}$ as a consequence of the increase in radius.
At the same time, the contraction of the convective core causes it to spin up to $10^{-2}$~s$^{-1}$.
H-burning starts at this point, which stops the contraction.
The slow expansion of the H-burning layers
translates into a slow decrease in the angular velocity of the core, towards $10^{-4}$~s$^{-1}$.
At that point, the envelope still only rotates at $\Omega\simeq10^{-8}-10^{-9}$~s$^{-1}$ ,
so that the rotation frequency in the core is four orders of magnitude higher than the frequency in the envelope. This is
in agreement with the rotating model of \cite{haemmerle2018b}.
This surface velocity is so low that the rotation period is of the order of a century.
Because the lifetime of the star is about $10^5\Ms/\dMac=10^5$~yr,
the surface only accomplishes some thousands of rotations during the entire stellar life.

The rate \dJ\ of increase in the angular momentum content of the star is shown in the fifth panel of Fig.~\ref{fig-ud} as the blue line.
We compare it with the angular momentum accretion rate \dJac, assumed Keplerian (Eq.~\ref{eq-jac}), shown as the grey line.
We deduce the required angular momentum loss \dJmag\ with Eq.~(\ref{eq-jdif}), and plot it as the red dashed line.
By comparing the curves, we see that the stellar angular momentum can only increase
by a few percent or less of the Keplerian angular momentum, in agreement with \cite{haemmerle2018b}.
Thus most of the accreted angular momentum must be removed by the magnetic field,
and the curve of \dJmag\ nearly matches that of \dJac.
Through the expansion of the stellar envelope, the Keplerian angular momentum grows much faster with time  than that of the star,
and thus a larger portion of the accreted angular momentum must be removed as the star grows in mass.

\subsection{Magnetic coupling}
\label{sec-mag2}

We estimated the Alfv\'en radius and the magnetic field at the stellar surface
according to the method described in Sect.~\ref{sec-mag1} (Eq.~\ref{eq-eta} -- \ref{eq-jmag}).
The two quantities are plotted in Fig.~\ref{fig-ud}.

The Alfv\'en radius (upper panel, red dashed line) evolves with the photospheric radius, keeping $r_A\sim100\,R$,
that is, $r_A\sim10^6\ \Rs\simeq5000$ AU.
This can be understood with the results of Sect.~\ref{sec-star2}: the magnetic field must remove essentially all the Keplerian angular momentum, meaning that $\dJmag\simeq\dJkep$.
This implies that (Eq.~\ref{eq-jac} and \ref{eq-jmag})
\begin{equation}
{2\over3}\dMwnd r_A^2\Omega_*=\dMac\sqrt{GMR}.
\end{equation}
Using Eq.~(\ref{eq-omgam}), we can express $r_A$ as a function of the Eddington factor:
\begin{equation}
\left({r_A\over R}\right)^2={1\over1-\Gamma}{\dMac\over\dMwnd}.
\end{equation}
With $\dMac=1$ \Mpy, $\dMwnd=0.001$ \Mpy\ (Sect.~\ref{sec-star1}), and $\Gamma=0.9-0.99$ (Fig.~\ref{fig-ud}, second panel), we obtain
\begin{equation}
\left({r_A\over R}\right)^2\sim10^4-10^5        \quad\Longrightarrow\quad       r_A\sim100-300\,R,
\end{equation}
in agreement with Fig.~\ref{fig-ud}.

The magnetic field required at the stellar surface is shown in the bottom panel of Fig.~\ref{fig-ud}.
During most of the evolution, the field must be $B_*\sim10^4$ G.
Stronger fields are required during the early phase (by about an order of magnitude),
but this reflects the choice of the initial model and its convergence towards the Hayashi limit.
This phase is extremely short (about a century), and the stellar mass remains $<100$ \Ms.
Here we focus on the supermassive range.
For $\sim10^5$ \Ms, the required magnetic field starts to increase,
but it remains weaker than $\sim3\times10^4$ G until the final collapse to the black hole.

\section{Discussion}
\label{sec-discuss}

\subsection{Angular momentum problem}

The model of this study was assumed to rotate at the \OG\ at each step of its evolution
and therefore corresponds to maximum rotation.
This is in contrast with the model previously published \citep{haemmerle2018b},
which accreted at a constant fraction of 1\% of the Keplerian angular momentum.
This allows us to quantify more precisely the constraints on the angular momentum accretion rate given by the \OG.
The fourth panel of Fig.~\ref{fig-ud} shows that the star can never accrete at the Keplerian limit.
In the early phase, the limit on \dJ\ is about 30\% of the Keplerian angular momentum,
but when the surface has converged to the Hayashi limit, the angular momentum cannot grow at more than 10\% of the Keplerian angular momentum.
For the rest of the evolution, \dJ\ is locked on a nearly constant value by the \OG,\ while the Keplerian angular momentum increases.
In the supermassive range, we need $\dJ<1\%$ \dJkep, in agreement with \cite{haemmerle2018b}.

The accretion disc might be sub-Keplerian, for instance in the presence of a pressure support.
In this case, the excess of angular momentum to be removed would be smaller than in the present case,
and thus weaker fields would be sufficient.
Our assumption of Keplerian accretion is the most conservative assumption.

The angular momentum problem is general to star formation \citep{spitzer1978,bodenheimer1995}.
While magnetic coupling with disc and winds provides a solution for low-mass stars,
additional mechanisms for extracting angular momentum, such as gravitational torques or viscosity, are required for massive stars.
All these mechanisms could play a role in the formation of SMSs \citep{hosokawa2016,takahashi2017,pandey2019}.
The mechanism addressed here can only maintain the star under the \OG.
Additional mechanisms are required to allow for accretion through a Keplerian disc.
Gravitational torques or viscosity are promising candidates because their efficiency increases with mass density:
if at a given stage an excess of angular momentum prevents accretion,
mass will accumulate in the disc until these processes are efficient enough for accretion to restart, in a dynamical time.
This implies fragmentation in the disc and episodic accretion \citep{sakurai2015,sakurai2016a}.
We note that the mechanism studied here does not rely on the accretion process and is not affected by accretion variability.

\subsection{Magnetic field}

The results of Sect.~\ref{sec-mag2} show that a magnetic field of $\sim10$ kG is required at the surface of the SMS.
This value is at the upper end of the observed range of stellar magnetic fields.
Magnetic fields of some kiloGauss are observed on $\sigma$ Ori E \citep{townsend2005}, HR 7355 \citep{rivinius2010}, or HR 5907 \citep{grunhut2012a}.
For the last star in particular, the inferred magnetic field is 10 -- 15 kG.

The magnetic properties of SMSs are not known.
We can estimate the surface magnetic field with the prescriptions derived from red giants.
The magnetic field is related to the Rossby number, which is defined as the ratio of the rotation period of the stellar envelope to the turnover time:
\begin{equation}
\Ro:={\tau_{\rm rot}\over\tau_{\rm tov}}.
\end{equation}
The relation is \citep{vidotto2014,auriere2015,privitera2016b}
\begin{equation}
\log B_*=-0.85\log\Ro+0.51,
\end{equation}
where $B_*$ is in Gauss.
For $\Omega_*\sim10^{-8}$ s$^{-1}$ (Fig.~\ref{fig-ud}), the rotation period is about 10 -- 100 yr.
The turnover time can be estimated with the formula \citep{rasio1996}
\begin{equation}
\tau_{\rm tov}^3={M_{\rm env}R_{\rm env}(R-R_{\rm env})\over3L},
\end{equation}
where $M_{\rm env}$ is the mass of the convective envelope, $R_{\rm env}$ is the radius at the base of the convective envelope,
and $L$ is the luminosity of the star.
For our models, this gives a turnover time of about a year.
Thus the Rossby number is $\Ro\sim10-100$ and the magnetic field is weaker than one Gauss (0.1 -- 1 G).
This is four orders of magnitude lower than the required values.
It shows that a convective dynamo cannot drive the required magnetic fields.

Highly magnetised stars such as $\sigma$ Ori E or HR 5907 do not have an extended convective envelope,
and their magnetic field is thought to originate from fossil fields.
Observations of intergalactic magnetic fields, which are thought to have primordial origins,
indicate a lower limit of $\sim10^{-16}$~G on megaparsec scales \citep{neronov2010}.
By flux conservation \citep{mouschovias1979},
the resulting field at the stellar surface ($\sim10^4$ \Rs, Fig.~\ref{fig-ud}) can be estimated by
\begin{equation}
B_*\gtrsim10^{-16}\rm\,G\,\left({1\,Mpc\over10^4\,\Rs}\right)^2\simeq2\,kG.
\end{equation}
Taken as a lower limit, this indicates that fossil fields from a primordial origin could easily reach the required 10 kG.

\section{Summary and conclusions}
\label{sec-out}

We have numerically derived the properties of Pop III SMSs at maximum rotation (\OG)
that accrete at the rates of atomically cooled haloes. We also analytically estimated the magnetic coupling with the winds
that is required for the star to rotate at such velocities while accreting from a Keplerian disc.

The derived stellar properties are consistent with those derived in previous studies
and refine the constraints on the angular momentum accretion rate.
We confirmed that the angular momentum of accreting SMSs cannot grow by more than $\sim1\%$ of the angular momentum from a Keplerian disc
during most of their accretion phase.
This in turn confirms that SMS formation by accretion requires mechanisms that are efficient enough
to remove more than 99\% of the Keplerian angular momentum.

We found that magnetic coupling between the stellar envelope and its mechanical winds
is efficient enough to remove this angular momentum excess
when the magnetic field at the stellar surface is of the order of 10 kG.
This value corresponds to the upper end of the observed range of stellar magnetic fields
and is consistent with fossil fields from a primordial origin.
For a sub-Keplerian accretion disc, weaker fields would be sufficient.
We obtained that the coupling holds up to a radial distance (Alfv\'en radius)
of about 100 times the photospheric radius, that is, $\sim5000$ AU.

We emphasize that this mechanism can only remove the angular momentum from the gas as it is accreted by the star.
Thus it cannot account for the angular momentum losses that are required for the gas to spiral in the Keplerian disc.
Additional mechanisms are necessary for this, such as viscosity \citep{takahashi2017},
gravitational torques \citep{hosokawa2016}, or large-scale magnetic fields \citep{pandey2019}.
The results of our study show that a magnetic field at the upper end of the observed stellar magnetic fields,
consistent with a primordial origin,
is able to remove the angular momentum excess that is accreted by the star, even if the disc remains Keplerian up to the stellar surface.

\begin{acknowledgements}
This work was sponsored by the Swiss National Science Foundation (project number 200020-172505).
\end{acknowledgements}

\bibliographystyle{aa}
\bibliography{bibliotheque}

\end{document}